\documentclass[a4paper]{article}

\usepackage{INTERSPEECH2022}

\usepackage{comment}
\usepackage[nolist,nohyperlinks]{acronym}
\usepackage{url}
\usepackage{booktabs}
\usepackage{amsmath}
\usepackage{textcomp}
\usepackage{graphicx}
\usepackage{subfig}
\usepackage{booktabs}
\usepackage[colorinlistoftodos,prependcaption]{todonotes}
\usepackage{xcolor}
\usepackage{pifont}
\usepackage{tabularx,colortbl}
\usepackage{cite}
\usepackage{amssymb}
\usepackage[]{hyperref}

\graphicspath{{./images/}}

\newcommand{\cmark}{\ding{51}}%
\newcommand{\xmark}{\ding{55}}%

\def\arrvline{\hfil\kern\arraycolsep\vline\kern-\arraycolsep\hfilneg}

\title{Joint domain adaptation and speech bandwidth extension using\\time-domain GANs for speaker verification}
\name{Saurabh Kataria$^{1,2}$, Jes\'us Villalba$^{1,2}$, Laureano Moro-Vel\'azquez$^{1}$, Najim Dehak$^{1,2}$}

\address{
$^{1}$Center for Language and Speech Processing, $^{2}$Human Language Technology Center of Excellence\\
Johns Hopkins University, Baltimore, MD, USA
  }
\email{\{skatari1,jvillal7,laureano,ndehak3\}@jhu.edu}

\begin{document}

\maketitle
 
\begin{abstract}
Speech systems developed for a particular choice of acoustic domain and sampling frequency do not translate easily to others.
The usual practice is to learn \emph{domain adaptation} and \emph{bandwidth extension} models independently. Contrary to this, we propose to learn both tasks together.
Particularly, we learn to map narrowband conversational telephone speech to wideband microphone speech.
We developed parallel and non-parallel learning solutions which utilize both paired and unpaired data.
First, we first discuss joint and disjoint training of multiple generative models for our tasks. %
Then, we propose a two-stage learning solution where we use a pre-trained domain adaptation system for pre-processing in bandwidth extension training.
We evaluated our schemes on a Speaker Verification downstream task. We used the JHU-MIT experimental setup for NIST SRE21, 
which comprises SRE16, SRE-CTS Superset and SRE21.
Our results provide the first evidence that learning both tasks is better than learning just one.
On SRE16, our best system achieves 22\% relative improvement in Equal Error Rate w.r.t. a \emph{direct} learning baseline and 8\% w.r.t. a strong bandwidth expansion system.
\end{abstract}

\section{Introduction}
\label{sec:intro}
Deep learning powered speech systems are becoming ubiquitous.
They are usually designed for a particular choice of sampling frequency and acoustic domain.
To deploy them in a different condition, we typically re-train them with relevant data or develop pre-processing solutions~\cite{garcia2019speaker}.
Commonly, two fundamental speech tasks, namely \emph{bandwidth extension} (BWE) and \emph{domain adaptation} (DA), are pursued for this.
Bandwidth extension refers to predicting missing higher-frequency (\emph{upperband}) information, while domain adaptation refers to the set of techniques aimed at improving robustness of model to choice of testing domain.
Both tasks are shown to be greatly beneficial \cite{li2015deep,nidadavolu2018investigation,nidadavolu2019investigation,nidadavolu2019cycle,nidadavolu2019low,kataria2021deep} but usually only one of both is employed.

Bandwidth extension has been pursued via supervised as well as unsupervised learning.
For supervised learning, \emph{mapping}~\cite{li2015deep} (deep regression) is typically used.
Common choice of generative models for supervised and unsupervised learning include Conditional Generative Adversarial Network (CGAN)~\cite{su2021bandwidth} and Cycle-consistent GAN (CycleGAN)~\cite{zhu2017unpaired,haws2019cyclegan} respectively.
Domain adaptation is an extensively studied problem as well.
\cite{wang2021multi} studies Multi-Source Domain Adaptation (MSDA) using Domain Adversarial Training (DAT)~\cite{ganin2016domain}.
In \cite{garcia2014unsupervised}, the authors perform unsupervised adaptation of a Probabilistic Linear Discriminant Analysis (PLDA) classifier.
CycleGANs are also popular for domain adaptation~\cite{nidadavolu2019cycle,nidadavolu2019low,kataria2021deep,nidadavolu2020unsupervised,nidadavolu2020single,hosseini2018multi}.

However, there are limited studies for joint learning of bandwidth extension or domain adaptation with other tasks.
Joint training of speech denoising and bandwidth extension is proposed in~\cite{seltzer2005robust} and~\cite{hou2020multi}.
In \cite{chen2019joint}, authors learn domain adaptation and discriminative feature learning.
In \cite{li2019joint}, authors learn class alignment in addition to domain alignment.
To our knowledge, there is no study for joint learning of bandwidth extension and domain adaptation.
We hypothesize that the gains from these two tasks can be enhanced by learning them together.
We propose to use them as pre-processing solutions to improve the downstream task of Automatic Speaker Verification (ASV) --which aims at determining if the speakers in two recordings are identical or not.
We are interested in Telephony Speaker Verification i.e. our test sets (except SRE21) are composed of only telephone signals with sampling frequency of 8KHz.
However, the training data for front-end (x-vector network~\cite{snyder2018x}) and back-end (PLDA~\cite{villalba2019state}) of ASV uses microphone speech, which has sampling frequency of 16 kHz and has mismatch w.r.t. telephone domain.
We propose to project all signals to richer space (16KHz wideband microphone speech) and follow a test-time extension and/or adaptation.

Our main contributions are: (1) We provide the first study for joint learning of \emph{domain adaptation} and \emph{bandwidth extension} using time-domain GANs and establish strong baselines on real test sets; (2) We provide the proof-of-concept that learning of these tasks can be synergistic while leveraging synthetic narrowband microphone data; (3) We provide formulations of joint training of supervised and unsupervised GANs and report significant improvements on SRE16 test set with our best scheme.

\begin{figure*}[t!]
    \centering
    \includegraphics[width=0.80\linewidth]{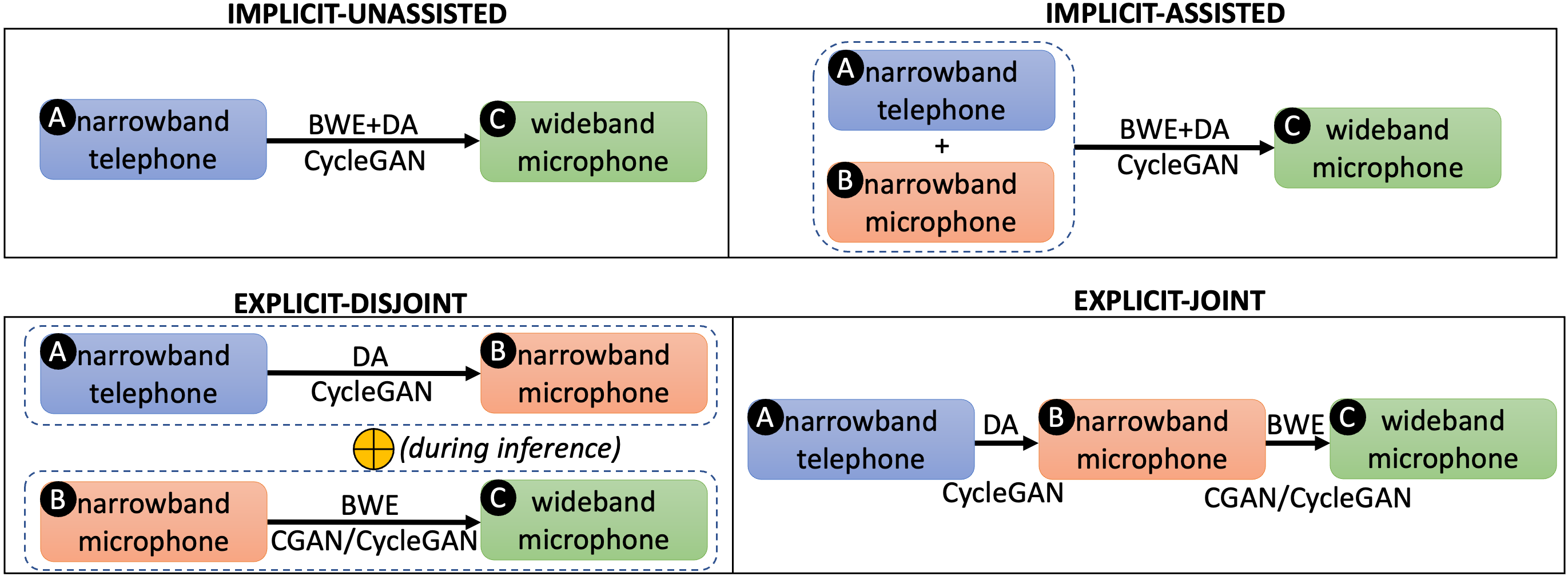}
    \vspace{-2mm}
    \caption{
    Illustration of four direct domain adaptation (DA) \& bandwidth extension (BWE) learning schemes.
    Each arrow denotes learning a mapping (marked with DA and/or BWE) from one domain to another.
    We also note where CGAN or CycleGAN can be used.
    }
    \label{fig:direct}
    \vspace{1.5em}
    \includegraphics[width=0.80\linewidth]{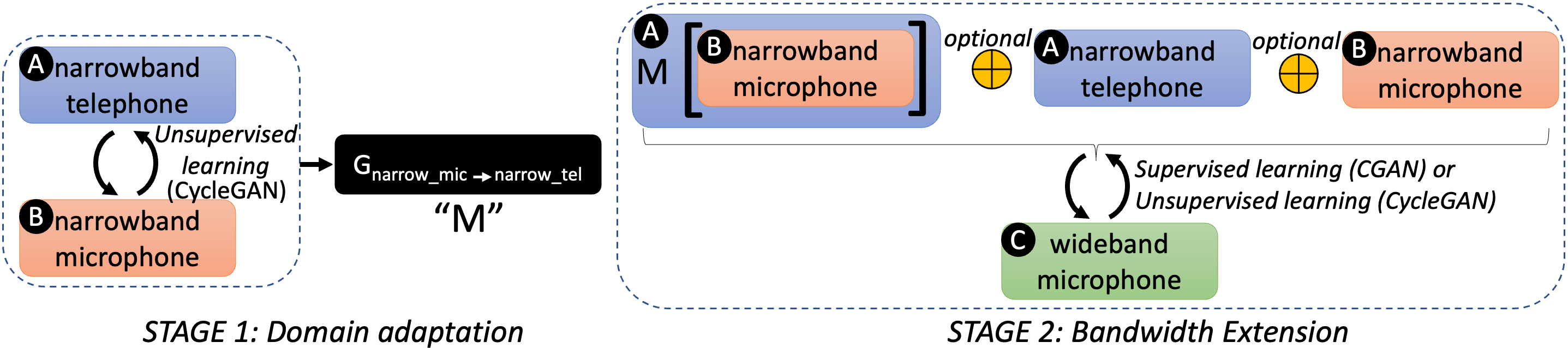}
    \vspace{-2mm}
    \caption{
    Illustration of indirect learning scheme.
    First stage learns adaptation $M=G_{\text{narrow\_mic}\rightarrow \text{narrow\_tel}}$.
    This is used to modify synthetic narrowband microphone data.
    Additionally, narrowband mic and tel data can be pooled in source domain for BWE training.
    }
    \label{fig:indirect}
    \vspace{-4mm}
\end{figure*}

\section{Supervised and unsupervised GANs}

Here, we describe the Generative Adversarial Network (GAN)~\cite{goodfellow2014generative} based models used in this work.
GANs learn to sample from true data distribution without access to class labels -- in our case, speaker labels.
Typically, they consist of two networks: generator and discriminator.
The discriminator learns to distinguish between real and fake/generated sample while the generator learns to ``fool'' the discriminator~\cite{goodfellow2014generative}.

\subsection{Conditional GAN}

CGAN is a supervised/\emph{parallel} learning GAN which learns mapping from one domain (say A) to another (say B) using paired data i.e. corresponding samples from two domains.
The generator is denoted by $\mathcal{G}_{A\rightarrow B}$ and discriminator for domain B is denoted by $\mathcal{D}_{B}$.
In addition to the typical adversarial loss of GAN ($\mathcal{L}_{\text{adv}}$), CGAN minimizes regression loss ($\mathcal{L}_{\text{sup}}$) between ground truth and predicted output, which is weighted by a hyper-parameter $\lambda_{\text{sup}}$.
The minimax formulation is:
\begin{align}
\label{eq:cgan}
    \min_{\mathcal{G}_{A\rightarrow B}} &\max_{\mathcal{D}_{B}} \mathcal{L}_{\text{CGAN},A,B} \hspace{0.5em} \text{, } \\
        \mathcal{L}_{\text{CGAN},A,B} &= \mathcal{L}_{\text{adv}} + \lambda_{\text{sup}}\mathcal{L}_{\text{sup}}.
\end{align}
$\mathcal{L}_{\text{adv}}$ (from Least-Squares GAN (LSGAN)~\cite{mao2017least}) is
\begin{equation}
\label{eq:adv}
    \mathbb{E}_{\mathbf{b}\sim p_{B}}[({\mathcal{D}}_{B}(\mathbf{b}))^2]\\
    + \mathbb{E}_{\mathbf{a}\sim p_{A}}[(1 - \mathcal{D}_{B}(\mathcal{G}_{A\rightarrow B}(\mathbf{a})))^2].
\end{equation}
The supervised loss $\mathcal{L}_{\text{sup}}$ or $\mathcal{L}_{\text{sup},A,B}$ is given by
\begin{equation}
    \mathcal{L}_{\text{sup},A,B} = \mathbb{E}_{\mathbf{a},\mathbf{b}\sim p_{A,B}} [\lVert\mathbf{b} - \mathcal{G}_{A\rightarrow B}(\mathbf{a})\rVert_2].
\end{equation}

\subsection{CycleGAN}
Cycle-consistent GAN is an unsupervised/\emph{non-parallel} learning GAN which learns mappings between two domains (say A and B) without pairing information between them.
It can be seen as two CGANs with additional cycle loss $\mathcal{L}_{\text{cyc}}$ and identity loss $\mathcal{L}_{\text{id}}$, which are weighted by hyper-parameters $\lambda_{\text{cyc}}$ and $\lambda_{\text{id}}$ respectively.
Cycle loss enforces semantic consistency in mapping while identity loss is a regularizer.
CycleGAN optimizes
\begin{align}
\label{eq:cyclegan}
    \min_{\mathcal{G}_{A\rightarrow B}, \mathcal{G}_{B\rightarrow A}} &\max_{\mathcal{D}_{A}, \mathcal{D}_{B}} \mathcal{L}_{\text{cyc-GAN},A,B} \hspace{0.5em} \text{, } \hspace{2em}\\
        \mathcal{L}_{\text{cyc-GAN},A,B} &= \mathcal{L}_{\text{adv},A\rightarrow B} + \mathcal{L}_{\text{adv},B\rightarrow A}\notag\\
        &+ \lambda_{\text{cyc}}\mathcal{L}_{\text{cyc}} + \lambda_{\text{id}}\mathcal{L}_{\text{id}}.
\end{align}
The expression for $\mathcal{L}_{\text{adv},A\rightarrow B}$ is identical to Eq.~\ref{eq:adv}, while $\mathcal{L}_{\text{adv},B\rightarrow A}$ is defined similarly as $\mathcal{L}_{\text{adv},A\rightarrow B}$ with A and B swapped.
Cycle and identity losses are given by
\begin{align}
    \mathcal{L}_{\text{cyc}} &= \mathbb{E}_{\mathbf{a}\sim p_{A},\mathbf{b}\sim p_{B}} [\lVert\mathbf{a} - \mathcal{G}_{B\rightarrow A}(\mathcal{G}_{A\rightarrow B}(\mathbf{a}))\rVert_1]\notag\\
    &+ \mathbb{E}_{\mathbf{a}\sim p_{A},\mathbf{b}\sim p_{B}} [\lVert\mathbf{b} - \mathcal{G}_{A\rightarrow B}(\mathcal{G}_{B\rightarrow A}(\mathbf{b}))\rVert_1]\\
    \mathcal{L}_{\text{id}} &= \mathbb{E}_{\mathbf{a}\sim p_{A}}[\lVert\mathbf{a} - \mathcal{G}_{B\rightarrow A}(\mathbf{a})\rVert_1] \notag \\
    &+ \mathbb{E}_{\mathbf{b}\sim p_{B}}[\lVert\mathbf{b} - \mathcal{G}_{A\rightarrow B}(\mathbf{b})\rVert_1]
\end{align}

\begin{table*}[t!]
\centering
\caption{
\label{tab:direct}
Results for direct learning schemes as EER / minDCF.
Bold numbers denotes best performance among two halves of the table.
}
\vspace{-4mm}
\resizebox{0.95\textwidth}{!}{%
\begingroup
\setlength{\tabcolsep}{3pt}
\begin{tabular}{@{}lrccc@{}}
\toprule
 \textbf{Name of direct scheme} & \textbf{Model} & \textbf{SRE16-YUE-eval40} & \textbf{SRE-CTS-superset-dev} & \textbf{SRE21-audio-eval} \\
\midrule
    \multicolumn{2}{@{}l}{\textbf{[\emph{narrow\_tel} = SRE-superset-train]}} &&&\\
   Implicit-unassisted [Baseline 1.1] & CycleGAN & 14.96 / 0.665 & 12.53 / 0.599 & 32.93 / 0.857 \\ %
   Implicit-assisted & CycleGAN & 13.47 / 0.637 & 9.41 / 0.428 & 30.35 / 0.796 \\ %
   Explicit-disjoint & CycleGAN $\oplus$ CGAN & \textbf{9.14} / \textbf{0.470} & 8.69 / 0.377 & \textbf{23.83} / \textbf{0.784} \\ %
   Explicit-disjoint & CycleGAN $\oplus$ CycleGAN & 9.68 / 0.482 & \textbf{8.63} / \textbf{0.370} & 24.72 / 0.788 \\ %
   Explicit-joint & CycleGAN $\oplus$ CGAN & 11.68 / 0.563 & 10.81 / 0.454 & 25.79 / 0.824 \\ %
   Explicit-joint & CycleGAN $\oplus$ CycleGAN & 25.45 / 0.875 & 20.44 / 0.795 & 39.82 / 0.941 \\ %
   \midrule
   \midrule
    \multicolumn{2}{@{}l}{\textbf{[\emph{narrow\_tel} = SRE-superset-train + SRE16-train]}} &&&\\
   Implicit-unassisted [Baseline 1.2] & CycleGAN & 8.71 / 0.452 & \textbf{5.37} / \textbf{0.240} & 21.34 / 0.743 \\ %
   Implicit-assisted & CycleGAN & \textbf{8.39} / \textbf{0.430} & 5.71 / \textbf{0.240} & \textbf{20.19} / \textbf{0.714} \\ %
   Explicit-disjoint & CycleGAN $\oplus$ CGAN & 11.43 / 0.552 & 9.20 / 0.393 & 23.91 / 0.796 \\ %
   Explicit-disjoint & CycleGAN $\oplus$ CycleGAN & 11.21 / 0.527 & 9.03 / 0.387 & 24.88 / 0.801 \\ %
   Explicit-joint & CycleGAN $\oplus$ CGAN & 12.22 / 0.572 & 10.18 / 0.459 & 23.26 / 0.782 \\ %
   Explicit-joint & CycleGAN $\oplus$ CycleGAN & 29.00 / 0.930 & 23.14 / 0.847 & 42.01 / 0.971 \\ %
\bottomrule
\end{tabular}
\endgroup
\vspace{-5mm}
}

\vspace{1.2em}
\caption{
\label{tab:indirect}
Results for indirect learning scheme as EER / minDCF.
$M$ is obtained from Stage 1 (CycleGAN) training (See Fig.~\ref{fig:indirect}).
}
\vspace{-4mm}
\resizebox{0.95\textwidth}{!}{%
\begingroup
\setlength{\tabcolsep}{3pt}
\begin{tabular}{@{}lrccc@{}}
\toprule
\textbf{Source data for Stage 2 model} & \textbf{Stage 2 Model} & \textbf{SRE16-YUE-eval40} & \textbf{SRE-CTS-superset-dev} & \textbf{SRE21-audio-eval} \\
\midrule

   \multicolumn{2}{@{}l}{\textbf{[\emph{narrow\_tel} = SRE-superset-train + SRE16-train]}} &&&\\
    \emph{narrow\_tel} [Baseline 1.2] & CycleGAN & 8.71 / 0.452 & 5.37 / 0.240 & 21.34 / 0.743 \\ %
    $M$(\emph{narrow\_mic}) & CGAN & 6.90 / 0.368 & 5.21 / 0.222 & 20.88 / 0.711 \\ %
    $M$(\emph{narrow\_mic}) + \emph{narrow\_mic} & CGAN & \textbf{6.75} / \textbf{0.358}
 & 5.37 / 0.221 & 20.09 / 0.688 \\ %
    $M$(\emph{narrow\_mic}) + \emph{narrow\_tel} & CycleGAN & 7.73 / 0.407
 & 5.37 / 0.227 & 21.65 / 0.706 \\ %
    $M$(\emph{narrow\_mic}) + \emph{narrow\_tel} + \emph{narrow\_mic} & CycleGAN & 7.51 / 0.392 & \textbf{5.00} / \textbf{0.202} & \textbf{18.77} / \textbf{0.679} \\ %
\bottomrule
\end{tabular}
\endgroup
}
\vspace{-3mm}
\end{table*}

\subsection{Joint training of CycleGAN and conditional GAN}
We propose to jointly train CycleGAN and CGAN to learn $A\leftrightarrow B$ and $B\rightarrow C$ mappings respectively, where C is the third domain of interest.
For this model (CycleGAN$\oplus$CGAN), we optimize:
\begin{equation}
\min_{\mathcal{G}_{A\rightarrow B}, \mathcal{G}_{B\rightarrow A}, \mathcal{G}_{B\rightarrow C}}  \max_{\mathcal{D}_{A}, \mathcal{D}_{B}, \mathcal{D}_{C}} \mathcal{L}_{\text{cyc-GAN,CGAN}}\hspace{0.5em} \text{,}
\end{equation}
\begin{align}
        &\mathcal{L}_{\text{cyc-GAN,CGAN}} = \mathcal{L}_{\text{cyc-GAN},A,B} + \mathcal{L}_{\text{CGAN},B,C}
        + \mathcal{L}_{\text{adv},B'\rightarrow C}\hspace{0.1em},\\
        &\mathcal{L}_{\text{adv},B'\rightarrow C} = \mathbb{E}_{\mathbf{a}\sim p_{A}}[(1 - \mathcal{D}_{C}(\mathcal{G}_{B\rightarrow C}(\mathcal{G}_{A\rightarrow B}(\mathbf{a}))))^2]
\end{align}
Here, $\mathcal{L}_{\text{cyc-GAN},A,B}$ and $\mathcal{L}_{\text{CGAN},B,C}$ are derived from Eq.~\ref{eq:cgan} \& \ref{eq:cyclegan}.
$\mathcal{L}_{\text{adv},B'\rightarrow C}$ is an additional adversarial loss which is used to tie together the training of CycleGAN and CGAN.

\subsection{Joint training of two CycleGANs}
We propose to train two CycleGANs which learns $A\leftrightarrow B$ and $B\leftrightarrow C$ mappings.
That is, we do not learn $A\rightarrow C$ and $C\rightarrow A$.
For this model (CycleGAN$\oplus$CycleGAN), we optimize:
\begin{equation}
\min_{\mathcal{G}_{A\rightarrow B}, \mathcal{G}_{B\rightarrow A}, \mathcal{G}_{B\rightarrow C},\mathcal{G}_{C\rightarrow B}}  \max_{\mathcal{D}_{A}, \mathcal{D}_{B}, \mathcal{D}_{C}} \mathcal{L}_{\text{cyc-GAN,cyc-GAN}}.%
\end{equation}
$\mathcal{L}_{\text{cyc-GAN,cyc-GAN}}$ is defined as
\begin{align}
    &\mathcal{L}_{\text{cyc-GAN,B,C}} + \mathcal{L}_{\text{cyc-GAN,A,B}} + 
    \mathcal{L}_{\text{adv},B'\rightarrow C} + \mathcal{L}_{\text{adv},B'\rightarrow A} \notag\\
    &+ \lambda_{\text{cyc}}(\mathcal{L}_{\text{cyc},B''} + \mathcal{L}_{\text{cyc},A'} + \mathcal{L}_{\text{cyc},C'}) \hspace{0.5em},
\end{align}
$\mathcal{L}_{\text{cyc},B''}$ is defined as:
\begin{align}
    &\mathbb{E}_{\mathbf{a}\sim p_{A}}[\lVert\mathcal{G}_{A\rightarrow B}(\mathbf{a}) - \mathcal{G}_{C\rightarrow B}(\mathcal{G}_{B\rightarrow C}(\mathcal{G}_{A\rightarrow B}(\mathbf{a})))\rVert_1 +\notag\\
    &\mathbb{E}_{\mathbf{c}\sim p_{C}}[\lVert\mathcal{G}_{C\rightarrow B}(\mathbf{c}) -
    \mathcal{G}_{A\rightarrow B}(\mathcal{G}_{B\rightarrow A}(\mathcal{G}_{C\rightarrow B}(\mathbf{c})))\rVert_1].
\end{align}
$\mathcal{L}_{\text{cyc},A'}$ is defined as:
\begin{align}
\label{eq:cycA'}
    \mathbb{E}_{\mathbf{a}\sim p_{A}}[\lVert\mathbf{a} -
    \mathcal{G}_{B\rightarrow C}(\mathcal{G}_{C\rightarrow B}(\mathcal{G}_{B\rightarrow C}(\mathcal{G}_{A\rightarrow B}(\mathbf{a}))))\rVert_1].
\end{align}
$\mathcal{L}_{\text{cyc},C'}$ is defined similarly to $\mathcal{L}_{\text{cyc},A'}$ with A and C swapped in Eq.~\ref{eq:cycA'}.
Note that we investigated additional adversarial terms in this model but obtained worse preliminary results.
\vspace{-2mm}
\section{Domain adaptation and bandwidth extension schemes}
There are three domains of interest: narrowband telephone (domain A or \emph{narrow\_tel}), narrowband microphone (domain B or \emph{narrow\_mic}), and wideband microphone (domain C or \emph{wide\_mic}).
Mapping from \emph{narrow\_tel} to \emph{narrow\_mic} ($A\rightarrow B$) refers to domain adaptation, while mapping from \emph{narrow\_mic} to \emph{wide\_mic} ($B\rightarrow C$) refers to bandwidth extension.
We are primarily interested in $A\rightarrow C$ but may learn more mappings depending on the scheme pursued. %
As per our experimental setup (Sec.~\ref{sec:exp}), (A,B) is unpaired data and hence, only unsupervised learning (via CycleGAN) is possible for $A\rightarrow B$.
(B,C) is paired data and thus, supervised learning can be used for $B\rightarrow C$.
Consequently, (A,C) represents unpaired data.
Note that it is possible to ignore pairing information in paired data and pursue unsupervised learning instead.
Our proposed schemes are categorized into \emph{direct} and \emph{indirect}.

\begin{table*}[t!]
\centering
\caption{
\label{tab:comp}
Comparison of proposed techniques with stronger baselines.
Supervised and unsupervised BWE is accomplished via CGAN and CycleGAN respectively (source data = narrow\_mic, target data = wide\_mic).
Bold numbers denote overall best results.
}
\vspace{-4mm}
\resizebox{0.85\textwidth}{!}{%
\begingroup
\setlength{\tabcolsep}{3pt}
\begin{tabular}{@{}ccccccccc@{}}
\toprule
 \textbf{BWE} & \textbf{DA} & \textbf{DA and/or BWE scheme} & \textbf{SRE16-YUE-eval40} & \textbf{SRE-CTS-superset-dev} & \textbf{SRE21-audio-eval} \\
\midrule
  \xmark & \xmark & - & 7.46 / 0.382 & 5.42 / 0.217 & 18.52 / 0.662 \\
  \cmark & \xmark & Supervised BWE & 7.35 / 0.376 & 4.90 / 0.205 & \textbf{17.02} / \textbf{0.648} \\ %
  \cmark & \xmark & Unsupervised BWE & 8.62 / 0.437 & 5.67 / 0.249 & 20.92 / 0.737 \\ %
  \xmark & \cmark & Unsupervised DA & 11.50 / 0.532 & 9.13 / 0.387 & 24.25 / 0.800 \\ %
  \midrule
  \cmark & \cmark & Direct (implicit-assisted) & 8.39 / 0.430 & 5.71 / 0.240 & 20.19 / 0.714 \\ %
  \cmark & \cmark & Indirect (semi-supervised) & \textbf{6.75} / \textbf{0.358} & 5.37 / 0.221 & 20.09 / 0.688 \\ %
  \cmark & \cmark & Indirect (unsupervised w/ addn. data) & 7.51 / 0.392 & 5.00 / \textbf{0.202} & 18.77 / 0.679 \\ %
\bottomrule
\end{tabular}
\endgroup
}
\vspace{-5mm}
\end{table*}
\vspace{-2mm}
\subsection{Direct learning schemes}
Direct learning schemes do not involve any pre-training stages and work with original (i.e. not pre-processed) data available from three domains (Fig.~\ref{fig:direct}).
They are further categorized into four schemes.
\emph{Implicit}-\emph{unassisted} is the simplest (yet most demanding) learning scheme which requires BWE and DA from A to C to be learned implicitly, i.e., mapping to and from domain B is not modeled and must be learned inherently without any assistance.
\emph{Implicit}-\emph{assisted} is a relaxed variant of \emph{implicit}-\emph{unassisted} where data from domain B is combined with domain A, thereby making learning easier. Both implicit schemes use CycleGAN since there is not
paired data between A and B.
\emph{Explicit} schemes use domain B data as intermediate targets to relax \emph{implicit} learning but at the cost of learning more mappings (two instead of one, see Fig.~\ref{fig:direct}).
\emph{Explicit}-\emph{disjoint} learns BWE and DA independently and uses the two learned mapping together during inference.
\emph{Explicit}-\emph{joint} learns the two tasks in the conventional multi-target joint learning manner. While the mapping from A to B has to be learned unsupervisedly by CycleGAN, the mapping from B to C can also be learned supervisedly with CGAN since we can create paired data.
\vspace{-2mm}
\subsection{Indirect (two-stage) learning scheme}
Indirect scheme involves two stage learning (Fig.~\ref{fig:indirect}).
In the first stage, domain adaptation is learned between narrowband data (\emph{narrow\_mic} and \emph{narrow\_tel}) using CycleGAN, which provides us $M$.
In the second stage, bandwidth extension is learned between \emph{wide\_mic} and \emph{narrow\_mic} (but projected to telephone domain via $M$).
Since \emph{narrow\_mic} is synthetically created (Sec.~\ref{sec:exp}) data, our motivation is to improve its utility via pre-processing.
This is inspired from our previous work~\cite{kataria2021deep} which showed that training data can be improved via domain adaptation pre-processing.
In this scheme, source domain data size for stage two can be increased by including telephone and unmodified narrowband microphone data as well for further gains.

\section{Experimental Setup}
\label{sec:exp}
For generators, we use off-the-shelf TasNet architecture which is a time-domain 1-D CNN~\cite{luo2019conv}.
We derive discriminator architectures from \cite{kong2020hifi}.
For CGAN, it is a single-period network with four initial channels, while for other models it is a multi-period network with eight initial channels.
X-vector architecture is Light ResNet-34~\cite{villalba2020advances,villalba2019jhu} and its input is 80-D log Mel-Frequency Filterbank (LMFB) features.
X-vector/speaker embedding dimension is 256 and PLDA dimension is 150.
We used smaller x-vector network (but with data augmentation) due to high computational complexity of x-vector extraction in doing multiple experiments.
The training data for \emph{narrow\_tel} is SRE superset~\cite{sadjadi2021nist} (SRE-superset-train) and SRE16 dev data (SRE16-train)~\cite{reynolds20172016,villalbajhu} which includes Tagalog and Cantonese (YUE) languages.
The training data for \emph{narrow\_mic} and \emph{wide\_mic} domains are VoxCeleb~\cite{nagrani2020voxceleb} downsampled to 8KHz, and VoxCeleb respectively.
Thus, (\emph{narrow\_tel}, \emph{narrow\_mic}) is unpaired data and (\emph{narrow\_mic}, \emph{wide\_mic}) is paired data.
For speaker verification, the training data for x-vector network and PLDA is combination of training data from three domains along with standard data augmentation with all MUSAN noises~\cite{snyder2015musan} and Aachen Room Impulse Response (RIR)~\cite{villalba2019state} reverberations.
We have three test sets: 40\% speakers of SRE16 Cantonese test set (SRE16-YUE-eval40), SRE Superset dev set, and SRE21 eval set.
The latter two consists of a wide range of languages.
SRE21 eval set also contains CTS-AFV (microphone vs telephone) speaker verification trials.
All signals at 8KHz (narrowband training data, all test sets) are resampled to 16KHz. %
Results are reported using ASV metrics (lower the better): Equal Error Rate (EER) (in \%) and minimum Decision Cost Function (minDCF) ($p_{\text{target}}=0.05$).
All GANs are trained with Alternating Gradient Descent~\cite{goodfellow2014generative} where the generators are updated at twice the frequency to that of the discriminators.
$(\lambda_{\text{sup}},\lambda_{\text{cyc}},\lambda_{\text{id}})=(0.1,10,5)$.
For CGAN, initial learning rates for generator and discriminator are 0.0004 and 0.0002, and batch size is 16.
Learning rates and batch size for CycleGAN are half as that of the CGAN.
Learning rates decay linearly with each step to 1e-8.
We use Adam~\cite{kingma2014adam} optimizer with betas=(0.5,0.999).

\vspace{-4mm}
\section{Results}
\subsection{Direct learning schemes}
To study the effect of using matching data in train and test, we conduct experiments with and without using SRE16-train.
Consider the upper-half of Table~\ref{tab:direct} which does not use SRE16-train in \emph{narrow\_tel}.
All schemes (except last) are able to outperform Baseline 1.1 (implicit-unassisted), which is the simplest scheme in terms of formulation.
Explicit-disjoint gets the best results.
Explicit-joint models give promising results but their performance lags due to two reasons: 1) CycleGAN$\oplus$CGAN and CycleGAN$\oplus$CycleGAN are computationally heavy models and we were unable to tune their hyper-parameters, 2) CycleGAN$\oplus$CycleGAN is fully unsupervised model which is difficult to train compared to supervised models.
Now, consider the upper-half of Table~\ref{tab:direct} which uses SRE16-train.
As expected, baseline (1.2) is much stronger.
Here, implicit-assisted achieves best results while explicit learning schemes achieve performance to their counterparts from the upper-half of table.
This indicates explicit schemes are promising and robust to \emph{narrow\_tel} choice. %
Note in explicit-disjoint schemes, replacing CGAN with CycleGAN retains performance, showing that paired and unpaired learning results can be close.
Also note that using time-domain GAN models bring significant challenges w.r.t. training stability and hyper-parameter choice.

\subsection{Indirect (two-stage) learning scheme}
Table~\ref{tab:indirect} presents the results for the indirect scheme.
Here, we include SRE16-train in \emph{narrow\_mic}.
Unsupervised learning of first stage (Fig.~\ref{fig:indirect}) gives us a pre-processor which can used during supervised training in second stage.
Thus, we are able to do supervised training while (indirectly) using unpaired (telephone) data.
CGAN models benefits from supervised learning and matching train-test data and gives best performance on SRE16 test set.
However, CycleGAN, which can leverage any amounts of unsupervised data, gets best performance on other two test sets using additional \emph{source} data.

\subsection{Comparison of all techniques}
We compare proposed techniques to stronger baselines in Table~\ref{tab:comp}.
The first row does not involve BWE or DA.
The second and third row does only BWE using CGAN and CycleGAN respectively.
In these experiments, we observe that BWE is essential to reduce EER/minDCF.
Fourth row does only unsupervised DA using CycleGAN and degrades performance.
This shows that, in our experimental setup, sampling frequency mismatch is a bigger challenge than domain mismatch.
Direct (implicit-assisted) is unsupervised method and is able to beat unsupervised baselines (row 3 and 4) but not supervised method (row 2).
It is important to note that supervised BWE (row 2) is a very strong baseline obtained via extensive tuning.
All other models in this work follow similar hyper-parameters as row 2, thereby giving sub-optimal results.
Indirect (semi-supervised) is from Table~\ref{tab:indirect} which uses CycleGAN (for pre-training) and CGAN, and hence is termed semi-supervised.
It gives best SRE16 test performance (8\% relative EER reduction w.r.t. row 2 and 22\% EER reduction w.r.t. baseline 1.2). %
However, it degrades performance on SRE21 which is retained by indirect model which uses additional data (last row).
We find SRE21 test set to be challenging. %
Finally, our schemes are able to beat three out of six metrics (w.r.t. strongest baseline), which demonstrates their potential.
Additional tuning, which is critical for GAN models, can bring in more improvements.
Since there is a mismatch in training objective of BWE/DA and ASV, speaker-identity preserving domain adaptation~\cite{kataria2021deep} can help.

\section{Conclusion}
We investigated joint learning of bandwidth extension and domain adaptation for improving telephony speaker verification.
We proposed two types of joint learning (direct and indirect) and provided schematic illustration.
Under direct learning scheme, we noted how synthetically created narrowband microphone data could be used in multiple ways, especially with joint training of multiple generative models like Conditional GAN and CycleGAN.
Under indirect learning scheme, we saw how multi stage learning can outperform strong baselines.
We also saw the critical role of choice of telephone training data.
In future, we can analyze the role of languages in the training and test sets, which seems to play a significant role.

\clearpage
\bibliographystyle{IEEEtran}
\bibliography{mybib}

\begin{thebibliography}{10}
\providecommand{\url}[1]{#1}
\csname url@samestyle\endcsname
\providecommand{\newblock}{\relax}
\providecommand{\bibinfo}[2]{#2}
\providecommand{\BIBentrySTDinterwordspacing}{\spaceskip=0pt\relax}
\providecommand{\BIBentryALTinterwordstretchfactor}{4}
\providecommand{\BIBentryALTinterwordspacing}{\spaceskip=\fontdimen2\font plus
\BIBentryALTinterwordstretchfactor\fontdimen3\font minus
  \fontdimen4\font\relax}
\providecommand{\BIBforeignlanguage}[2]{{%
\expandafter\ifx\csname l@#1\endcsname\relax
\typeout{** WARNING: IEEEtran.bst: No hyphenation pattern has been}%
\typeout{** loaded for the language `#1'. Using the pattern for}%
\typeout{** the default language instead.}%
\else
\language=\csname l@#1\endcsname
\fi
#2}}
\providecommand{\BIBdecl}{\relax}
\BIBdecl

\bibitem{garcia2019speaker}
P.~Garc{\'\i}a, J.~Villalba, H.~Bredin, J.~Du, D.~Castan, A.~Cristia,
  L.~Bullock, L.~Guo, K.~Okabe, P.~S. Nidadavolu \emph{et~al.}, ``Speaker
  detection in the wild: Lessons learned from jsalt 2019,'' \emph{arXiv
  preprint arXiv:1912.00938}, 2019.

\bibitem{li2015deep}
K.~Li and C.-H. Lee, ``A deep neural network approach to speech bandwidth
  expansion,'' in \emph{2015 IEEE International Conference on Acoustics, Speech
  and Signal Processing (ICASSP)}.\hskip 1em plus 0.5em minus 0.4em\relax IEEE,
  2015, pp. 4395--4399.

\bibitem{nidadavolu2018investigation}
P.~S. Nidadavolu, C.-I. Lai, J.~Villalba, and N.~Dehak, ``Investigation on
  bandwidth extension for speaker recognition.'' in \emph{INTERSPEECH}, 2018,
  pp. 1111--1115.

\bibitem{nidadavolu2019investigation}
P.~S. Nidadavolu, V.~Iglesias, J.~Villalba, and N.~Dehak, ``Investigation on
  neural bandwidth extension of telephone speech for improved speaker
  recognition,'' in \emph{ICASSP 2019-2019 IEEE International Conference on
  Acoustics, Speech and Signal Processing (ICASSP)}.\hskip 1em plus 0.5em minus
  0.4em\relax IEEE, 2019, pp. 6111--6115.

\bibitem{nidadavolu2019cycle}
P.~S. Nidadavolu, J.~Villalba, and N.~Dehak, ``Cycle-gans for domain adaptation
  of acoustic features for speaker recognition,'' in \emph{ICASSP 2019-2019
  IEEE International Conference on Acoustics, Speech and Signal Processing
  (ICASSP)}.\hskip 1em plus 0.5em minus 0.4em\relax IEEE, 2019, pp. 6206--6210.

\bibitem{nidadavolu2019low}
P.~S. Nidadavolu, S.~Kataria, J.~Villalba, and N.~Dehak, ``Low-resource domain
  adaptation for speaker recognition using cycle-gans,'' in \emph{2019 IEEE
  Automatic Speech Recognition and Understanding Workshop (ASRU)}.\hskip 1em
  plus 0.5em minus 0.4em\relax IEEE, 2019, pp. 710--717.

\bibitem{kataria2021deep}
S.~Kataria, J.~Villalba, P.~{\.Z}elasko, L.~Moro-Vel{\'a}zquez, and N.~Dehak,
  ``Deep feature cyclegans: Speaker identity preserving non-parallel
  microphone-telephone domain adaptation for speaker verification,''
  \emph{arXiv preprint arXiv:2104.01433}, 2021.

\bibitem{su2021bandwidth}
J.~Su, Y.~Wang, A.~Finkelstein, and Z.~Jin, ``Bandwidth extension is all you
  need,'' in \emph{ICASSP 2021-2021 IEEE International Conference on Acoustics,
  Speech and Signal Processing (ICASSP)}.\hskip 1em plus 0.5em minus
  0.4em\relax IEEE, 2021, pp. 696--700.

\bibitem{zhu2017unpaired}
J.-Y. Zhu, T.~Park, P.~Isola, and A.~A. Efros, ``Unpaired image-to-image
  translation using cycle-consistent adversarial networks,'' in
  \emph{Proceedings of the IEEE international conference on computer vision},
  2017, pp. 2223--2232.

\bibitem{haws2019cyclegan}
D.~Haws and X.~Cui, ``Cyclegan bandwidth extension acoustic modeling for
  automatic speech recognition,'' in \emph{ICASSP 2019-2019 IEEE International
  Conference on Acoustics, Speech and Signal Processing (ICASSP)}.\hskip 1em
  plus 0.5em minus 0.4em\relax IEEE, 2019, pp. 6780--6784.

\bibitem{wang2021multi}
Z.~Wang and J.~H. Hansen, ``Multi-source domain adaptation for text-independent
  forensic speaker verification,'' \emph{IEEE/ACM Transactions on Audio,
  Speech, and Language Processing}, 2021.

\bibitem{ganin2016domain}
Y.~Ganin, E.~Ustinova, H.~Ajakan, P.~Germain, H.~Larochelle, F.~Laviolette,
  M.~Marchand, and V.~Lempitsky, ``Domain-adversarial training of neural
  networks,'' \emph{The journal of machine learning research}, vol.~17, no.~1,
  pp. 2096--2030, 2016.

\bibitem{garcia2014unsupervised}
D.~Garcia-Romero, A.~McCree, S.~Shum, N.~Brummer, and C.~Vaquero,
  ``Unsupervised domain adaptation for i-vector speaker recognition,'' in
  \emph{Proceedings of Odyssey: The Speaker and Language Recognition Workshop},
  vol.~8, 2014.

\bibitem{nidadavolu2020unsupervised}
P.~S. Nidadavolu, S.~Kataria, J.~Villalba, P.~Garcia-Perera, and N.~Dehak,
  ``Unsupervised feature enhancement for speaker verification,'' in
  \emph{ICASSP 2020-2020 IEEE International Conference on Acoustics, Speech and
  Signal Processing (ICASSP)}.\hskip 1em plus 0.5em minus 0.4em\relax IEEE,
  2020, pp. 7599--7603.

\bibitem{nidadavolu2020single}
P.~S. Nidadavolu, S.~Kataria, P.~Garc{\'\i}a-Perera, J.~Villalba, and N.~Dehak,
  ``Single channel far field feature enhancement for speaker verification in
  the wild,'' \emph{arXiv preprint arXiv:2005.08331}, 2020.

\bibitem{hosseini2018multi}
E.~Hosseini-Asl, Y.~Zhou, C.~Xiong, and R.~Socher, ``A multi-discriminator
  cyclegan for unsupervised non-parallel speech domain adaptation,''
  \emph{arXiv preprint arXiv:1804.00522}, 2018.

\bibitem{seltzer2005robust}
M.~L. Seltzer, A.~Acero, and J.~Droppo, ``Robust bandwidth extension of
  noise-corrupted narrowband speech,'' in \emph{Ninth European conference on
  speech communication and technology}.\hskip 1em plus 0.5em minus 0.4em\relax
  Citeseer, 2005.

\bibitem{hou2020multi}
N.~Hou, C.~Xu, J.~T. Zhou, E.~S. Chng, and H.~Li, ``Multi-task learning for
  end-to-end noise-robust bandwidth extension.'' in \emph{INTERSPEECH}, 2020,
  pp. 4069--4073.

\bibitem{chen2019joint}
C.~Chen, Z.~Chen, B.~Jiang, and X.~Jin, ``Joint domain alignment and
  discriminative feature learning for unsupervised deep domain adaptation,'' in
  \emph{Proceedings of the AAAI conference on artificial intelligence},
  vol.~33, no.~01, 2019, pp. 3296--3303.

\bibitem{li2019joint}
S.~Li, C.~H. Liu, B.~Xie, L.~Su, Z.~Ding, and G.~Huang, ``Joint adversarial
  domain adaptation,'' in \emph{Proceedings of the 27th ACM International
  Conference on Multimedia}, 2019, pp. 729--737.

\bibitem{snyder2018x}
D.~Snyder, D.~Garcia-Romero, G.~Sell, D.~Povey, and S.~Khudanpur, ``X-vectors:
  Robust dnn embeddings for speaker recognition,'' in \emph{2018 IEEE
  International Conference on Acoustics, Speech and Signal Processing
  (ICASSP)}.\hskip 1em plus 0.5em minus 0.4em\relax IEEE, 2018, pp. 5329--5333.

\bibitem{villalba2019state}
J.~Villalba, N.~Chen, D.~Snyder, D.~Garcia-Romero, A.~McCree, G.~Sell,
  J.~Borgstrom, F.~Richardson, S.~Shon, F.~Grondin \emph{et~al.},
  ``State-of-the-art speaker recognition for telephone and video speech: The
  jhu-mit submission for nist sre18.'' in \emph{Interspeech}, 2019, pp.
  1488--1492.

\bibitem{goodfellow2014generative}
I.~J. Goodfellow, J.~Pouget-Abadie, M.~Mirza, B.~Xu, D.~Warde-Farley, S.~Ozair,
  A.~Courville, and Y.~Bengio, ``Generative adversarial networks,'' \emph{arXiv
  preprint arXiv:1406.2661}, 2014.

\bibitem{mao2017least}
X.~Mao, Q.~Li, H.~Xie, R.~Y. Lau, Z.~Wang, and S.~Paul~Smolley, ``Least squares
  generative adversarial networks,'' in \emph{Proceedings of the IEEE
  international conference on computer vision}, 2017, pp. 2794--2802.

\bibitem{luo2019conv}
Y.~Luo and N.~Mesgarani, ``Conv-tasnet: Surpassing ideal time--frequency
  magnitude masking for speech separation,'' \emph{IEEE/ACM transactions on
  audio, speech, and language processing}, vol.~27, no.~8, pp. 1256--1266,
  2019.

\bibitem{kong2020hifi}
J.~Kong, J.~Kim, and J.~Bae, ``Hifi-gan: Generative adversarial networks for
  efficient and high fidelity speech synthesis,'' \emph{Advances in Neural
  Information Processing Systems}, vol.~33, pp. 17\,022--17\,033, 2020.

\bibitem{villalba2020advances}
J.~Villalba, D.~Garcia-Romero, N.~Chen, G.~Sell, J.~Borgstrom, A.~McCree,
  L.~Garcia-Perera, S.~Kataria, P.~S. Nidadavolu, P.~A. Torres-Carrasquillo
  \emph{et~al.}, ``Advances in speaker recognition for telephone and
  audio-visual data: the jhu-mit submission for nist sre19,'' in
  \emph{Proceedings of Odyssey}, 2020.

\bibitem{villalba2019jhu}
J.~Villalba, D.~Garcia-Romero, N.~Chen, G.~Sell, J.~Borgstrom, A.~McCree,
  D.~Snyder, S.~Kataria, P.~Garc{\i}a-Perera, F.~Richardson \emph{et~al.},
  ``The jhu-mit system description for nist sre19 av,'' in \emph{NIST SRE19
  Workshop}, 2019.

\bibitem{sadjadi2021nist}
S.~O. Sadjadi, ``Nist sre cts superset: A large-scale dataset for telephony
  speaker recognition,'' \emph{arXiv preprint arXiv:2108.07118}, 2021.

\bibitem{reynolds20172016}
D.~Reynolds, E.~Singer, S.~O. Sadjadi, T.~Kheyrkhah, A.~Tong, C.~Greenberg,
  L.~Mason, and J.~Hernandez-Cordero, ``The 2016 nist speaker recognition
  evaluation,'' MIT Lincoln Laboratory Lexington United States, Tech. Rep.,
  2017.

\bibitem{villalbajhu}
J.~Villalba, J.~Borgstrom, S.~Kataria, J.~Cho, P.~A. Torres-Carrasquillo, and
  N.~Dehak, ``The jhu-mit system description for nist sre20 cts challenge.''

\bibitem{nagrani2020voxceleb}
A.~Nagrani, J.~S. Chung, W.~Xie, and A.~Zisserman, ``Voxceleb: Large-scale
  speaker verification in the wild,'' \emph{Computer Speech \& Language},
  vol.~60, p. 101027, 2020.

\bibitem{snyder2015musan}
D.~Snyder, G.~Chen, and D.~Povey, ``Musan: A music, speech, and noise corpus,''
  \emph{arXiv preprint arXiv:1510.08484}, 2015.

\bibitem{kingma2014adam}
D.~P. Kingma and J.~Ba, ``Adam: A method for stochastic optimization,''
  \emph{arXiv preprint arXiv:1412.6980}, 2014.

\end{thebibliography}
\end{document}